\PassOptionsToPackage{table}{xcolor}
\documentclass[conference]{IEEEtran}
\usepackage{amsthm, amsmath}
\usepackage{float}
\usepackage{array}
\usepackage{booktabs}
\usepackage{graphicx}
\usepackage{subcaption}
\usepackage[justification=centering]{caption}
\usepackage[font=small]{caption}
\usepackage[hyphens]{url}
\usepackage[bookmarks=false]{hyperref}
\usepackage{comment}
\hypersetup{breaklinks=true}
\usepackage{cite}
\usepackage[export]{adjustbox}
\usepackage{tabularx}
\usepackage[utf8]{inputenc}
\usepackage{lscape} 
\usepackage{color}
\usepackage{afterpage}
\usepackage{multirow}
\usepackage{enumitem}
\usepackage{array}
\usepackage{soul}
\usepackage{subcaption}
\usepackage[font=small]{caption}
\newcommand{\PreserveBackslash}[1]{\let\temp=\\#1\let\\=\temp}
\newcolumntype{L}[1]{>{\raggedright\let\newline\\\arraybackslash\hspace{0pt}}m{#1}}
\newcolumntype{C}[1]{>{\centering\let\newline\\\arraybackslash\hspace{0pt}}m{#1}}
\newcolumntype{R}[1]{>{\raggedleft\let\newline\\\arraybackslash\hspace{0pt}}m{#1}}
\usepackage{tikz}
\usepackage{longtable}
\usepackage{tablefootnote}
\usepackage[para,flushleft]{threeparttable}
\usepackage{adjustbox}
\usepackage{array}
\usepackage{amssymb}
\usepackage{siunitx} 
\usepackage[T1]{fontenc}

\newcolumntype{M}[2]{%
    >{\adjustbox{angle=#1,lap=\width-(#2)}\bgroup}%
    l%
    <{\egroup}%
}
\newcommand*\rot[2]{\multicolumn{1}{M{#1}{#2}}}

\usepackage[table]{xcolor}

\newcommand{\benchname}{Koios}

\begin{document}


\title{
\benchname: A Deep Learning Benchmark Suite for FPGA Architecture and CAD Research
\vspace{-0.4cm}
}

\author{
\footnotesize
\IEEEauthorblockN{
Aman Arora$^1$, 
Andrew Boutros$^2$,
Daniel Rauch$^1$,
Aishwarya Rajen$^1$,
Aatman Borda$^1$,
Seyed Alireza Damghani$^3$,\\
Samidh Mehta$^1$,
Sangram Kate$^1$,
Pragnesh Patel$^1$,
Kenneth B. Kent$^3$,
Vaughn Betz$^2$,
Lizy K. John$^1$
}

\IEEEauthorblockA{
$^1$\textit{The University of Texas at Austin} ~~~~~
$^2$\textit{University of Toronto \& Vector Institute for AI} ~~~~~
$^3$\textit{University of New Brunswick}
\\
E-mail: aman.kbm@utexas.edu
}
\vspace{-0.4cm}
}




\begin{figure*}[!t]
\begin{large}
\textcopyright 2021 IEEE. Personal use of this material is permitted.  Permission from IEEE must be obtained for all other uses, in any current or future media, including reprinting/republishing this material for advertising or promotional purposes, creating new collective works, for resale or redistribution to servers or lists, or reuse of any copyrighted component of this work in other works.
\newline
\newline
This work has been accepted at the 2021 31st International Conference on Field-Programmable Logic and Applications (FPL) and will appear in the proceedings and on the IEEE website on/around August 31, 2021.
\newline
\newline

\end{large}
\end{figure*}
\pagebreak

\maketitle

\begin{abstract}
With the prevalence of deep learning (DL) in many applications, researchers are investigating different ways of optimizing FPGA architecture and CAD to achieve better quality-of-results (QoR) on DL-based workloads.
In this optimization process, benchmark circuits are an essential component; the QoR achieved on a set of benchmarks is the main driver for architecture and CAD design choices.
However, current academic benchmark suites are inadequate, as they do not capture any designs from the DL domain.
This work presents a new suite of DL acceleration benchmark circuits for FPGA architecture and CAD research, called Koios.
This suite of 19 circuits covers a wide variety of accelerated neural networks, design sizes, implementation styles, abstraction levels, and numerical precisions.
These designs are larger, more data parallel, more heterogeneous, more deeply pipelined, and utilize more FPGA architectural features compared to existing open-source benchmarks.
This enables researchers to pin-point architectural inefficiencies for this class of workloads and optimize CAD tools on more realistic benchmarks that stress the CAD algorithms in different ways. 
In this paper, we describe the designs in our benchmark suite, present results of running them through the Verilog-to-Routing (VTR) flow using a recent FPGA architecture model, and identify key insights from the resulting metrics. 
On average, our benchmarks have 3.7$\times$ more netlist primitives, 1.8$\times$ and 4.7$\times$ higher DSP and BRAM densities, and 1.7$\times$ higher frequency with 1.9$\times$ more near-critical paths compared to the widely-used VTR suite. 
Finally, we present two example case studies showing how architectural exploration for DL-optimized FPGAs can be performed using our new benchmark suite. 

\end{abstract}

\IEEEpeerreviewmaketitle


\section{Introduction} \label{section:introduction}

With compute and data intensive deep learning (DL) becoming a major component of many applications, specialized hardware acceleration of such workloads has become a commonplace. 
More recently, field-programmable gate arrays (FPGAs) have been shown to deliver state-of-the-art performance when accelerating different DL workloads because of their massive parallelism, flexibility and energy efficiency \cite{hall2020hpipe, boutrosbeyond}. 
With new DL use cases emerging faster than ever, FPGAs are also starting to adapt. 
This includes the emergence of DL-optimized FPGA fabrics \cite{langhammer2021stratix}, the integration of FPGAs with specialized DL accelerators \cite{nurvitadhi2019compete, ahmad2019xilinx}, and also tuning FPGA CAD tools to the properties of these workloads \cite{zhang2019frequency}. 

In general, the development of novel FPGA architectures and CAD algorithms depends mainly on a versatile framework that consists of three main components: (1) a set of benchmarks written in a hardware description language or synthesized using high-level synthesis, (2) an architecture model that captures the organization of FPGA blocks and routing architecture as well as area/timing/power models from circuit-level implementations, and (3) a CAD flow that synthesizes the given benchmarks then implements them on a given FPGA architecture \cite{fpga_arch_evolution}.
Although most research efforts in the FPGA community are focused on architecture and CAD, benchmarks actually play a crucial role in this flow.
The quality-of-results (QoR) achieved on a specific set of benchmarks is the main driver for architecture and CAD design choices.
As a result, it is essential that these benchmarks capture the markets and application domains targeted by the candidate FPGA architecture.
Using an unrepresentative set of benchmarks means optimizing for the wrong targets. 

Among the existing open-source benchmark suites, which we will discuss in a later section, none of them focus on (or even capture any) benchmarks from the increasingly important DL domain.
Therefore, it becomes very tedious to evaluate architecture and CAD optimizations for DL-targeted FPGAs, since researchers have to first implement their own benchmarks.
This limits any research efforts in this direction to only individual isolated ones, and makes it virtually impossible to have meaningful comparisons between different ideas across the FPGA research community.
Our work addresses this by presenting \benchname\footnote{\textit{Koios} (also written as \textit{Coeus}) is the Titan of intelligence in Greek mythology. Unlike the Titan benchmarks, our suite focuses on deep learning.}, an open-source benchmark suite of DL acceleration benchmark circuits for FPGA architecture and CAD research.
This suite consists of 19 benchmarks that capture a wide variety of accelerated neural networks, design sizes, numerical precisions, and circuit characteristics.
To maximize the utility of these benchmarks, we made them compatible with the Verilog-to-Routing (VTR) flow \cite{vtr8}, which is arguably the most widely-used FPGA architecture and CAD research framework. Researchers can use these benchmarks seamlessly with VTR and with minor modifications, can even use them with other toolchains.

Koios benchmarks are representative of modern DL workloads; many of them are re-created from prior works and some are replicas of industrial architectures. In addition to being more pipelined and DSP/BRAM intensive, these benchmarks have higher usage of structures like wide busses, large reduction trees, hard block cascades and large fanouts. This makes Koios benchmarks much better suited for DL-targeted FPGA architecture exploration than any non-DL benchmark suite.

All the benchmarks along with the FPGA architecture we used for our experiments in this paper are open-sourced as a part of VTR\footnote{\href{https://tinyurl.com/vtrkoios}{https://tinyurl.com/vtrkoios}}.
In this paper, we make the following contributions:
\begin{itemize}[noitemsep,topsep=0pt,leftmargin=2\labelsep]
    \item Introduce the \benchname\ benchmarks and describe the different characteristics of the constituent designs. 
    \item Present the results of running our benchmarks through VTR using an FPGA architecture description file that we develop to capture complex DSP features typical of recent FPGAs.
    \item Compare circuit statistics to those of the VTR benchmarks to highlight the added value of our new suite.
    \item Describe two example case studies that use these benchmarks to explore architectural optimizations for DL.
\end{itemize} 
\section{Related Work} \label{section:related_work}

\subsection{FPGA Benchmark Suites}
There are several benchmark suites that were used by FPGA architecture and CAD researchers throughout the past three decades. 
The classic MCNC20 benchmarks \cite{mcnc_benchmarks} are extremely small and simple designs that do not use any FPGA hard blocks.
Therefore, they do not represent modern FPGA use-cases and are rarely used for architecture or CAD studies nowadays. 
The twenty largest circuits from this suite (often referred to as the Toronto20 \cite{vpr}) are provided in the input format consumed by the Versatile Place and Route (VPR) tool suite. 
The UMass RCG HDL Benchmark Collection \cite{umass_rcg} has larger designs mostly representing DSP applications. 
However, this suite does not target an open-source FPGA framework. 
The Groundhog benchmarks \cite{groundhog_benchmarks} are shown to work with academic toolflows and are targeted towards evaluation of power consumption of FPGAs for mobile computing applications. 
ERCBench \cite{ercbench} is another suite consisting of hybrid hardware/software applications. 
The designs in this suite represent designs from multimedia, wireless communications and cryptography. They do not contain DL benchmarks, and  do not work with academic FPGA tools. 

VTR \cite{vtr8} has a suite of benchmarks as well. 
These VTR benchmarks vary from small ($321$ netlist primitives) to medium-sized designs ($165,809$ primitives) and they capture a multitude of applications like image processing, soft processors and arithmetic.
The Titan benchmark suite \cite{titan} contains modern heterogeneous large designs ($90$K to $1.8$M netlist primitives). 
However, they target a hybrid CAD flow that is architecture-specific as logic synthesis is performed using the Intel Quartus flow only for the Stratix IV architecture. 
In contrast to all existing suites, \benchname\ is the only one that provides large, heterogeneous, architecture-agnostic benchmarks that work with a completely open-source flow such as VTR, and focuses on the increasingly important DL domain.


\subsection{DL-Optimized FPGAs}
Recently, FPGA vendors have released products with many DL-targeted features to cater to the ever-growing demands of DL workloads. 
For example, the Xilinx Versal ACAP \cite{xilinx_versal_ai} added specialized vector processors for DL acceleration, and Intel's Stratix 10 NX devices integrated in-fabric AI tensor blocks \cite{langhammer2021stratix}.
In addition, the announced Achronix Speedster7t FPGAs \cite{achronix_mlp} will have embedded machine learning processor (MLP) blocks that tightly couple memory and compute for DL, and the FlexLogix nnMAX \cite{flexlogix_nnmax} inference IP also contains tiles with hardened convolution logic. 
For their architecture exploration, FPGA vendors typically use proprietary customer designs or internal benchmarks that are not accessible to the research community.

There have also been a number of academic research proposals for optimizing FPGA architectures for DL.
Eldafrawy et al. \cite{logic_block_mohd} proposed several enhancements to the logic block architecture to pack more arithmetic bits or add a shadow multiplier in them for improved DL performance. 
They used simple multiplier/MAC and 4$\times$4 matrix multiplication microbenchmarks to evaluate their proposed ideas. 
In \cite{embrace_div_dsp, pir_dsp}, the authors explored enhancing DSP blocks by efficiently supporting low precision multiplications. 
For these studies, the authors design their own benchmarks to evaluate their ideas. 
Arora et al. \cite{tensor_slice_paper} also proposed adding Tensor slices in FPGAs.
Again, they use their own designs, a TPU-like overlay and several microbenchmarks, for their evaluation. 
We believe that an open-source  benchmark suite is needed to create a common ground for evaluating and comparing such FPGA architectural enhancements for DL. 





\begin{table*}[htbp]
  \rowcolors{2}{gray!15}{white}
  \caption{The \benchname\ Benchmarks (in decreasing order of number of netlist primitives)}
  \begin{threeparttable}
    \begin{tabular}  {l@{\hspace{1\tabcolsep}} l@{\hspace{1\tabcolsep}} l@{\hspace{1\tabcolsep}} l@{\hspace{1.5\tabcolsep}} l@{\hspace{1\tabcolsep}}l
    @{\hspace{1\tabcolsep}}l@{\hspace{0.5\tabcolsep}}l@{\hspace{0.5\tabcolsep}}l@{\hspace{0.5\tabcolsep}}l@{\hspace{0.5\tabcolsep}}l@{\hspace{0.5\tabcolsep}}l
    @{\hspace{1\tabcolsep}}l@{\hspace{1\tabcolsep}}l}
    \toprule
    
    \rot{0}{0.5em}{\textbf{Benchmark}} & \rot{0}{0.5em}{\textbf{Description}} & \rot{45}{0.5em}{\textbf{Implementation}} & \rot{45}{0.5em}{\textbf{Network}} & \rot{45}{0.5em}{\textbf{Precision}} & \rot{45}{0.5em}{\textbf{Acc. Paradigm}} & \rot{45}{0.5em}{\textbf{2D Systolic}} & \rot{45}{0.5em}{\textbf{Winograd/FFT}} & \rot{45}{0.5em}{\textbf{Reduction}} & \rot{45}{0.5em}{\textbf{Buffers}} & \rot{45}{0.5em}{\textbf{DSP usage}} & \rot{45}{0.5em}{\textbf{Cent. buffers}} & \rot{45}{0.5em}{\textbf{Based on}} & \rot{0}{0.5em}{\textbf{Other Properties}} \\
    \toprule
    clstm\_like (S/M/L) & CLSTM-like accelerator & RTL   & RNN   & int18 & Overlay &    &\checkmark    &    & \checkmark \tnote{3}    &\checkmark    &    & \cite{c_lstm_paper} & Circular compression \\
    dla\_like (S/M) & Intel-DLA-like accelerator & RTL   & CNN\tnote{2} & int8/16 & Overlay &&\checkmark    &    & \checkmark \tnote{3}    & \checkmark \tnote{4}   &\checkmark    & \cite{intel_dla_orig}\cite{benchmark_arch} & Daisy chain \\
    lstm  & LSTM engine & RTL   & RNN   & int16 & Layer &    &    &\checkmark    &\checkmark    &\checkmark    &    &       & Streaming dataflow \\
    tpu\_like (S/M) & Google-TPU-v1-like accelerator & RTL   & Any\tnote{1} \tnote{2} & int8  & Overlay &\checkmark    &    &    &\checkmark    &\checkmark    &\checkmark    & \cite{tpu_v1_paper} & APB interface \\
    bnn   & 4-layer binary neural network & HLS & MLP\tnote{1} & binary & Custom &    &    &    &    &\checkmark    &    & \cite{hls4ml} \cite{bnn_hls4ml_paper} & int16 act/norm \\
    tiny\_darknet\_like & Accelerator for Tiny Darknet & HLS & CNN\tnote{1} \tnote{2} & fp16  & Custom &    &    &    & \checkmark \tnote{3}     &\checkmark   &    & \cite{tiny_darknet} & Fused layer pairs \\
    gemm\_layer & Matrix multiplication engine & RTL   & MLP   & bfloat16 & Layer & \checkmark     &    &    &    &\checkmark    &\checkmark    &       & AXI interface \\
    attention\_layer & Transformer self-attention layer & RTL   & RNN   & int16 & Layer &    &    &\checkmark    & \checkmark \tnote{3}    &\checkmark    &    & \cite{attention_paper} & GEMV based \\
    conv\_layer & GEMM based convolution & RTL   & CNN   & int16 & Layer &\checkmark    &    &    &\checkmark    &\checkmark    &\checkmark    &       & 3x3 filters \\
    spmv  & Sparse matrix vector multiplication & RTL   & MLP   & int8  & Layer &    &    &    &\checkmark    &\checkmark    &\checkmark    & \cite{spmv_cisr} \cite{xilinx_gemx} & COO sparsity enc. \\
    robot\_rl & Robot+maze application  & RTL   & RL    & int8/16/32 & Custom &    &    &    &\checkmark    &\checkmark    &\checkmark    & \cite{reinforcement_learning_italy} \cite{reinforcement_learning_brazil} & Q-learning algo \\
    reduction\_layer & Add/max/min reduction tree & RTL   & Any   & int16 & Layer &    &    &\checkmark    &\checkmark    &    &\checkmark    &       & Reduces 128 inputs \\
    softmax & Softmax classification layer & RTL   & Any   & fp16  & Layer &    &    &\checkmark    &    &\checkmark    &    & \cite{softmax_asap20} & LUT based exp/log \\
    conv\_layer\_hls & Sliding window convolution & HLS & CNN   & fp16  & Layer &    &    &    &\checkmark    &\checkmark    &    &       & 1x1 filters  \\
    eltwise\_layer & Matrix elementwise add/sub/mult & RTL   & Any   & bfloat16 & Layer &    &    &    &\checkmark    &\checkmark    &\checkmark    &       & Broadcast heavy \\
    
    \bottomrule
    \end{tabular}%
    \begin{tablenotes}
    \item[1] Has Normalization layer\item[2] Has pooling layer     \item[3] Uses double buffering     \item[4] Has DSP cascade chains
  \end{tablenotes}
    \end{threeparttable}
  \label{table:airsuite_benchmarks}%
\end{table*}%

\section{The \benchname\ Benchmark Suite} \label{section:benchmark_description}

Our collection of benchmark designs in the \benchname\ suite come from a multitude of applications within the DL domain. 
They cover a wide variety of different design sizes, implementation styles, target neural networks, acceleration paradigms, numerical precisions, and circuit properties as summarized by the overview in Table \ref{table:airsuite_benchmarks}, and detailed in this section.

\begin{itemize}[noitemsep,topsep=0pt,leftmargin=2\labelsep]
    \item \textbf{Design Size}: 
    The smallest design has $11,519$ netlist primitives while the largest has $1,085,877$.
    Any latch, gate or hard block resulting from logic synthesis counts as a netlist primitive.
    Some benchmarks, such as \texttt{clstm\_like, dla\_like, tpu\_like}, have multiple size variants (i.e. small, medium, large). 
    In these cases, the size indicates the parallelism factor used in the design.
    Bigger designs create a more challenging optimization problem for the CAD tools, while smaller ones have faster compilation time suitable for early-stage architecture and CAD experiments.
    
    \item \textbf{Implementation Style}: 
    Although all the designs in the benchmark suite are provided to users in the form of Verilog HDL implementations, some were originally implemented in RTL while others were automatically generated from higher level language descriptions using high-level synthesis (HLS) tools.
    HLS-generated designs typically have specific design characteristics that are not generally seen in hand-coded RTL designs, such as widely distributed control signals and complex state machines.
    
    \item \textbf{Target Neural Network}: 
    Our benchmarks cover all major classes of neural networks.
    These include: multi-layer perceptrons (MLPs), convolutional neural networks (CNNs), recurrent neural networks (RNNs), and reinforcement learning (RL).
    These different classes have different compute and memory requirements, which reflects on the resource breakdown and routing patterns of their corresponding benchmark circuits.
    Some designs are also generic and can be used to accelerate any type of network.
    
    \item \textbf{Acceleration Paradigm}: 
    FPGAs are used for acceleration of DL workloads in different ways. 
    One way is to design a flexible software-programmable overlay architecture that can execute different DL models without the need to reprogram the FPGA with a new bitstream similar to the Microsoft Brainwave \cite{microsoft_brainwave} architecture.
    These designs tend to have instruction decoders and more complicated control logic to enable this level of flexibility.
    In other cases, a custom network-specific architecture is mapped to an FPGA to maximize efficiency similar to the approach used in \cite{hall2020hpipe}.
    The control logic of these circuits is usually hard-coded and implemented as relatively simple state machines.
    Another approach is to implement layer-specific accelerators that are invoked by software running on the host CPU.
    These circuits are mostly streaming-style datapaths with simple or even no control paths.
    Our benchmark suite contains designs from all three acceleration paradigms.

    \item \textbf{Numerical Precisions}: 
    One of the main advantages of using FPGAs to accelerate DL workloads is the ability to design hardware for custom numerical precisions, which is a commonly used technique in accelerating DL workloads \cite{darvish2020pushing}. 
    The designs in our suite use various precisions, including: binary (\texttt{bin}), different fixed point types \texttt{int8/16/32}, brain floating point (\texttt{bfloat16}) \cite{bfloat16}, and IEEE half-precision floating point (\texttt{fp16}).
    The diversity in the benchmarks' numerical precisions is useful for exploring new reconfigurable DSP block architectures and different hard arithmetic circuitry.
   
    \item \textbf{Circuit Properties}: 
    Our benchmarks have varying circuit styles that can potentially exercise different components of the CAD tools in different ways.
    For example, regular structures like systolic arrays can be used for optimizing placement algorithms, large reduction trees can form local routing congestions that stress the routing algorithms, long cascades of hard blocks impose harder placement constraints, etc.
    The benchmarks are also highly heterogeneous (i.e. use different types of FPGA resources) with varying degrees as will be discussed in Section \ref{section:benchmark_results}.
\end{itemize}

These benchmarks are implemented and compiled together in this suite with the intention to be used for FPGA architecture exploration and CAD tool optimization.
They aim to accurately capture all these different circuit structures and compositions, but should not be expected to be deployed as standalone functional systems. 
We are confident that these circuits are structurally correct and tried to verify their high-level functionality to the best of our ability.
However, full functional verification on many different test cases is out of the scope of this work.

\section{Methodology} \label{section:experimental_setup}

\subsection{Ensuring VTR Compatibility}
The designs in the benchmark suite are implemented and tested first using commercial FPGA tools from Xilinx and Intel for ease of development and debugging. 
Then, we performed several modifications to these designs to ensure their compatibility with the VTR flow. 
VTR uses Odin II, an academic open source synthesis tool, as its conventional front-end.
To work around the Verilog support limitations of Odin II and at the same time maintain the conventional fully open-source VTR flow, we implemented several scripts to help automate the process of replacing unsupported Verilog constructs (e.g. \texttt{signed}, \texttt{integer} variables, \texttt{generate} for loops, unpacked arrays, etc.) with alternative/unrolled Verilog constructs that are supported by Odin II.    
In addition, vendor-specific and architecture-specific IP cores (e.g. floating point adders and multipliers, RAM macros) were replaced with ones that are compatible with VTR and the FPGA architecture file used for our experiments. 
This process was especially challenging for the designs generated from HLS tools which tend to be non-human-readable in many cases. 
Several improvements to the language coverage and reported error messages of Odin II are continuously being implemented to mitigate such challenges for future research efforts.

\subsection{Experimental Setup}
We use the most-recent VTR 8.0 version \cite{vtr8} for all our experiments in this paper. 
While running VTR, we provide an SDC (Synopsys Design Constraints) file in which the target clock frequency is set to 0 (i.e. VTR will optimize the design for maximum clock frequency). 
We also disable timing analysis for paths to/from the FPGA IOs. 
For all experiments, we run VTR with auto layout enabled (meaning the grid size expands based on the resources required by the design), the default timing-driven routing option with a maximum of 150 routing iterations, and a fixed channel width of 300 wires.
All reported results are the average of runs using 3 different seeds.
For experiments in which we report VTR flow runtime and peak memory usage, we use an Intel Xeon CPU E5-2430 running at 2.5 GHz with 64 GB of memory.

One of the main motivations of this work is to compare various properties of our \benchname\ benchmarks with other existing non-DL-targeted benchmarks that are commonly used to drive FPGA architecture and CAD research. 
The most relevant suite for comparison is the VTR benchmark suite, because these are compatible with the same fully open source VTR flow.
Other existing suites are either too small and do not represent realistic modern use cases of FPGAs or depend partially on commercial CAD tools.
For these comparative experiments, we only use the VTR benchmarks with more than $10,000$ netlist primitives, which is a common practice in CAD-related studies \cite{elgammal2020learn}. 
Designs smaller than that are not representative of realistic benchmarks and they cannot be used to derive any reliable conclusions.


\subsection{FPGA Architecture Description}  
\label{section:fpga_arch_used}

We develop a new FPGA architecture description file to capture some relevant features of modern FPGAs.
This architecture description file will be open sourced along with the benchmark suite.
The delays and areas of all the FPGA blocks, including the DSP tiles, are obtained from COFFE \cite{coffe2} using a 22nm technology node from PTM \cite{asu_ptm}.
The circuits in this architecture are optimized for area-delay product which leads to relatively higher delays compared to performance-optimized commercial FPGAs such as the Arria 10 family.
The rest of this subsection describes the details of the FPGA architecture that we develop and use for all our experiments.

\subsubsection{Floorplan}
The FPGA contains columns of logic blocks, DSPs and block RAMs (BRAMs). 
Both DSP and BRAM columns repeat every 16 columns and are interleaved such that every 8th column is a DSP or a BRAM.
The DSP and BRAM tiles are 4 and 2 rows high, respectively.
IO pads are arranged along the perimeter of the FPGA. 

\subsubsection{Routing Architecture}
The architecture uses unidirectional routing with wire segments of length 4 (260 out of 300 wires) and length 16 (40 out of 300 wires). 
The length 16 wires do not directly connect to block pins and are only accessible from the length 4 wires. 
Switches appear after every 4 blocks on the length 16 wires. 
The switch blocks use a custom switching pattern based on the Stratix-IV-like architecture used in the Titan flow \cite{titan}. 
The input and output flexibility of connection blocks are set to 0.15 and 0.1, respectively.

\subsubsection{Logic Blocks}
Each logic block (LB) contains 10 basic logic elements (BLEs) similar to that in the Intel Stratix-10-like architecture from  \cite{logic_block_mohd}. 
Each block has 60 input pins, 40 output pins, and a 50\% sparsely populated local input crossbar. 
Each BLE has a 6-input LUT which can be fractured into two 5-input LUTs. The BLE also has 2 flip-flops and 2 bits of arithmetic with dedicated carry chains between LBs. Each BLE has 8 inputs and 4 optionally registered outputs. 

\subsubsection{DSP Slices}
This architecture has a complex DSP block that supports most of the operating modes in the state-of-the-art Intel Agilex DSP block \cite{intel_agilex}. 
Multiple fixed point (9x9, 18x19, 27x27) and floating point (IEEE 32-bit (\texttt{fp32}), IEEE 16-bit (\texttt{fp16}) and Brain floating point (\texttt{bfloat16})) precisions are supported.
In addition, the DSP block has dedicated output chains for cascading several DSP blocks in the same column for efficient dot product structures.

\subsubsection{BRAMs}
BRAM blocks have a capacity of 20 Kilobits and have registered inputs and outputs. True and simple dual port modes are supported.  
In the simple dual port mode, a BRAM can be configured as: 512$\times$40, 1024$\times$20 and 2048$\times$10, while in true dual port mode it can be configured only as: 1024$\times$20 and 2048$\times$10.
The delays and areas of a BRAM block are obtained by interpolation between the values obtained from COFFE for a 16 Kilobit BRAM and a 32 Kilobit BRAM.

Some benchmarks in \benchname\ use advanced DSP features that are available in this FPGA architecture by instantiating DSP macros to implement native \texttt{fp16} multiplications or use the hard dedicated chains. 
These modes are architecture-specific; however, users can simply replace the macro instantiations in our benchmarks with their equivalents for different architectures.
In addition, we also include alternative versions of the benchmarks (using \texttt{\textasciigrave ifdef..\textasciigrave endif}) implementing the same functionality with behavioral Verilog that is automatically mapped to the FPGA soft logic when an architecture without the required macro definitions is used. 

Koios benchmarks can be used to explore FPGA architectural modifications involving adding new hard blocks to FPGAs, similar to some recent DL-optimized FPGAs \cite{langhammer2021stratix} \cite{tensor_slice_paper}. This can be done by: (1) modifying the synthesis engine to extract specific patterns from the Verilog design and map them to the new blocks, or (2) modifying the benchmarks to instantiate these new blocks (defined in the VTR architecture file).

\begin{table*}[htbp]
  \centering
  \caption{VTR results of the \benchname\ benchmarks.}
  \rowcolors{2}{gray!15}{white}
    \begin{threeparttable}
    \begin{tabular}{L{2cm} R{1.5cm} R{1cm} R{1cm} R{1cm} R{1cm} R{1cm} R{1cm} R{1.5cm} R{1cm} R{1.2cm}}
    \toprule
    \textbf{Benchmark} & \textbf{Netlist} ~~~ \textbf{Primitives} &	\textbf{Logic Depth}	& \textbf{Used IOs} &	\textbf{Used LBs} &	\textbf{Used DSPs} &	\textbf{Used BRAMs} &	\textbf{Max. Freq.} &	\textbf{Routed Wirelength} &	\textbf{Elapsed Time} &	\textbf{Peak Memory}  \\
    \toprule
    
    clstm\_like (L) & 1,085,877 & 3     & 1,159  & 25,995 & \textbf{962}   & 1,161  & 110.2 & 5,534,505 & 1,171.4 & 12,658.4 \\
    
    clstm\_like (M) & 745,829 & 3     & 871   & 17,641 & \textbf{662}   & 784   & 115.4 & 3,612,133 & 560.6 & 8,691.1 \\
    
    dla\_like (M) & 609,180 & 5     & 411   & 11,359 & 400   & \textbf{1,008}  & 125.9 & 3,349,783 & 260.7 & 6,009.7 \\
    
    clstm\_like (S) & 405,776 & 3     & 583   & 9,309  & \textbf{362}   & 407   & 127.6 & 1,744,947 & 152.8 & 4,679.1 \\
    
    dla\_like (S) & 269,040 & 5     & 207   & 5,545  & 128   & \textbf{828}   & 147.7 & 1,475,558 & 86.1 & 4,304.7 \\
    
    lstm  & 249,841 & 7     & 36    & 6,626  & \textbf{610}   & 305   & 121.6 & 1,828,974 & 308.2 & 5,892.8 \\
    
    tpu\_like (M) & 244,884 & 5     & 1,188  & 4,255  & \textbf{1,064}  & 26    & 98.62 & 2,412,297 & 156.1 & 9,163.1 \\
    
    bnn   & 204,601 & 3     & 382   & \textbf{5,695}  & 63    & 0     & 126.8 & 1,233,543 & 20.9 & 2,153.1 \\
    
    tiny\_darknet\_like & 154,096 & 6     & 46    & 7,417  & 106   & \textbf{3,978}  & 63.9 & 3,033,846 & 571.1 & 16,253.5 \\
    
    tpu\_like (S) & 67,086 & 5     & 644   & 1,134  & \textbf{276}   & 14    & 124.8 & 579,437 & 31.9  & 2,507.5 \\
    
    gemm\_layer & 64,792 & 4     & 1,779  & 1,989  & \textbf{200}   & 0     & 308.1 & 717,412 & 25.4 & 1,982.2 \\
    
    attention\_layer & 45,342 & 7     & 1,074  & 1,248  & \textbf{105}   & 161   & 132.2 & 370,030 & 16.7 & 1,152.3 \\
    
    conv\_layer & 45,039 & 4     & 156   & 1,185  & \textbf{84}    & 56    & 166.1 & 293,011 & 9.4  & 876.3 \\
    
    spmv  & 28,505 & 6     & 19    & 885   & 32    & \textbf{257}   & 167.9 & 275,500 & 14.5 & 1,492.8 \\
    
    robot\_rl & 28,080 & 15    & 387   & 1,324  & 18    & \textbf{96}    & 83.6 & 228,378 & 9.1  & 549.5 \\
    
    reduction\_layer & 18,323 & 6     & 54    & 805   & 0     & \textbf{52}    & 141.7 & 183,739 & 2.2  & 363.2 \\
    
    softmax & 13,189 & 10    & 552   & 518   & \textbf{53}    & 0     & 112.2 & 127,704 & 2.5  & 513.3 \\
    
    conv\_layer\_hls & 12,093 & 3     & \textbf{3,299}  & 1,715  & 12    & 21    & 164.7 & 112,362 & 19.2 & 8,929.1 \\
    
    eltwise\_layer & 11,519 & 4     & 249   & 348   & \textbf{48}    & 72    & 174.9 & 170,857 & 2.1  & 480.9 \\
    \bottomrule
    \end{tabular}%
    \begin{tablenotes}
    \item Frequency is in MHz, Routed Wirelength is in units of length 1 segments, Elapsed Time is in minutes, and Peak Memory is in MBs.
    \end{tablenotes}
    \end{threeparttable}
    \vspace{-0.2cm}
  \label{tab:benchmark_properties}%
\end{table*}%
\section{Benchmark Results} \label{section:benchmark_results}

\subsection{Properties of \benchname\ benchmarks}

Table \ref{tab:benchmark_properties} shows the main VTR results for the \benchname\ benchmarks when running them with the FPGA architecture described in Section \ref{section:fpga_arch_used}.

The results show that these designs, with sizes ranging from $11$K to $1$M netlist primitives, are deeply-pipelined with 12 out of the 19 benchmarks having critical paths with 5 or less logic levels on them.
The benchmarks are also highly diverse in heterogeneity, with varying circuit compositions between soft logic, DSPs, and BRAMs.
For example, some designs do not utilize any BRAMs since they either implement only the workload datapath (e.g. \texttt{gemm\_layer} and \texttt{softmax}) or use distributed registers for storage (e.g. \texttt{bnn}).
On the other hand, there are other BRAM-intensive designs such as \texttt{tiny\_darknet\_like} with close to $4,000$ BRAMs utilized.
Similarly with DSPs, there are some designs that use very few or no DSPs (e.g. \texttt{bnn} and \texttt{reduction\_layer}) as they mostly implement other non-multiplication operations in DL workloads such as pop-count or max/min/add reduction.
Other designs are DSP-intensive (e.g. large \texttt{clstm\_like} and medium \texttt{tpu\_like}) with around $1,000$ DSP blocks.
Table \ref{tab:benchmark_properties} also shows that different types of resources are the grid-size limiting factor for different benchmarks in our suite.
The majority of the designs are bound by hard blocks, as indicated by the bold entries in the table, which emphasizes that these benchmarks can be useful for exploring new DSP and BRAM architectures.

\begin{figure}[t!]
    \centering
    \includegraphics[width=1\linewidth]{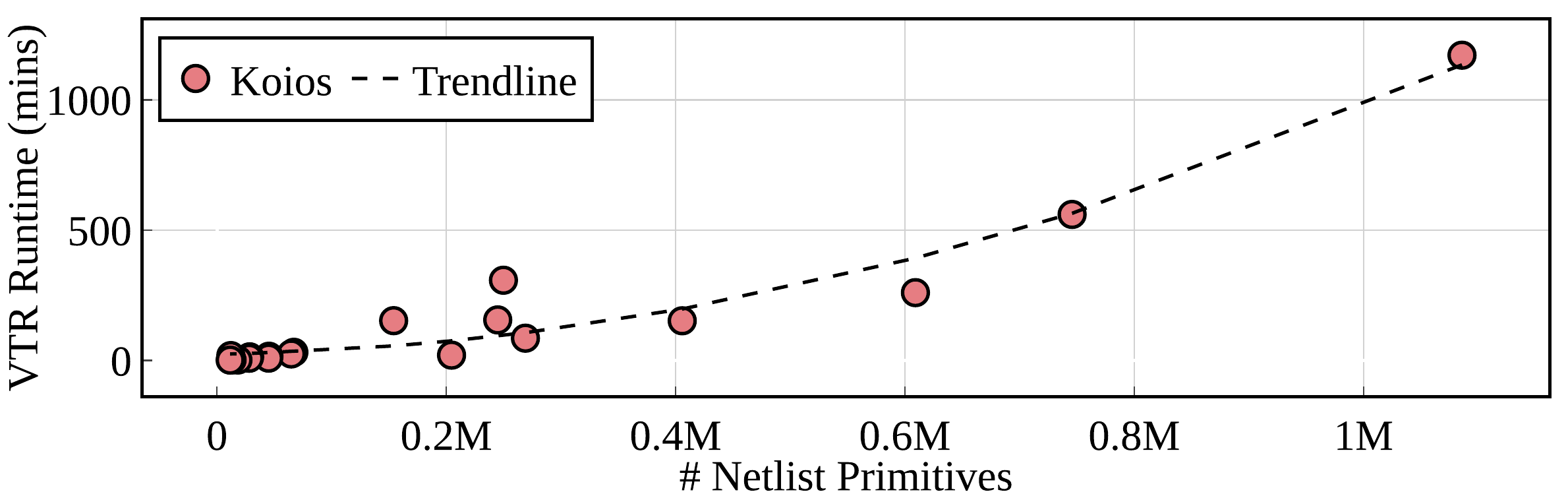}
    \caption{VTR runtime for the \benchname\ benchmarks.}
    \label{fig:runtime}
    \vspace{-0.2cm}
\end{figure}

Most of the designs in the \benchname\ suite can achieve reasonably high operating frequencies up to 308 MHz and an average of 137 MHz. The FPGA architecture used for our experiments is not very fast. The delays in the architecture are based on area-delay-optimized PTM models (with raw delays similar to 40 nm Stratix-IV). Changing the delays of FPGA resources to those typical of a high-speed ($\leq $14 nm) device would increase the frequency by $>$2$\times$. 
The \texttt{tiny\_darknet\_like} design is a clear outlier with a frequency of 63.9 MHz since the grid size required to implement this circuit was significantly expanded due to the large number of BRAMs needed.
This resulted in some very long paths between BRAMs and soft logic flip-flops (FFs).
The total routed wirelength of the benchmarks are largely correlated with the circuit size and ranges from $171$K up to $5.5$M units of length 1 wire segments.
Fig. \ref{fig:runtime} plots the VTR flow runtime for each of the \benchname\ benchmarks as listed in Table \ref{tab:benchmark_properties}.
It shows that the runtime grows quadratically with the number of netlist primitives in the circuits.

\subsection{Comparison to the VTR Benchmarks}

\begin{figure}[t!]
    \centering
    \includegraphics[width=1\linewidth]{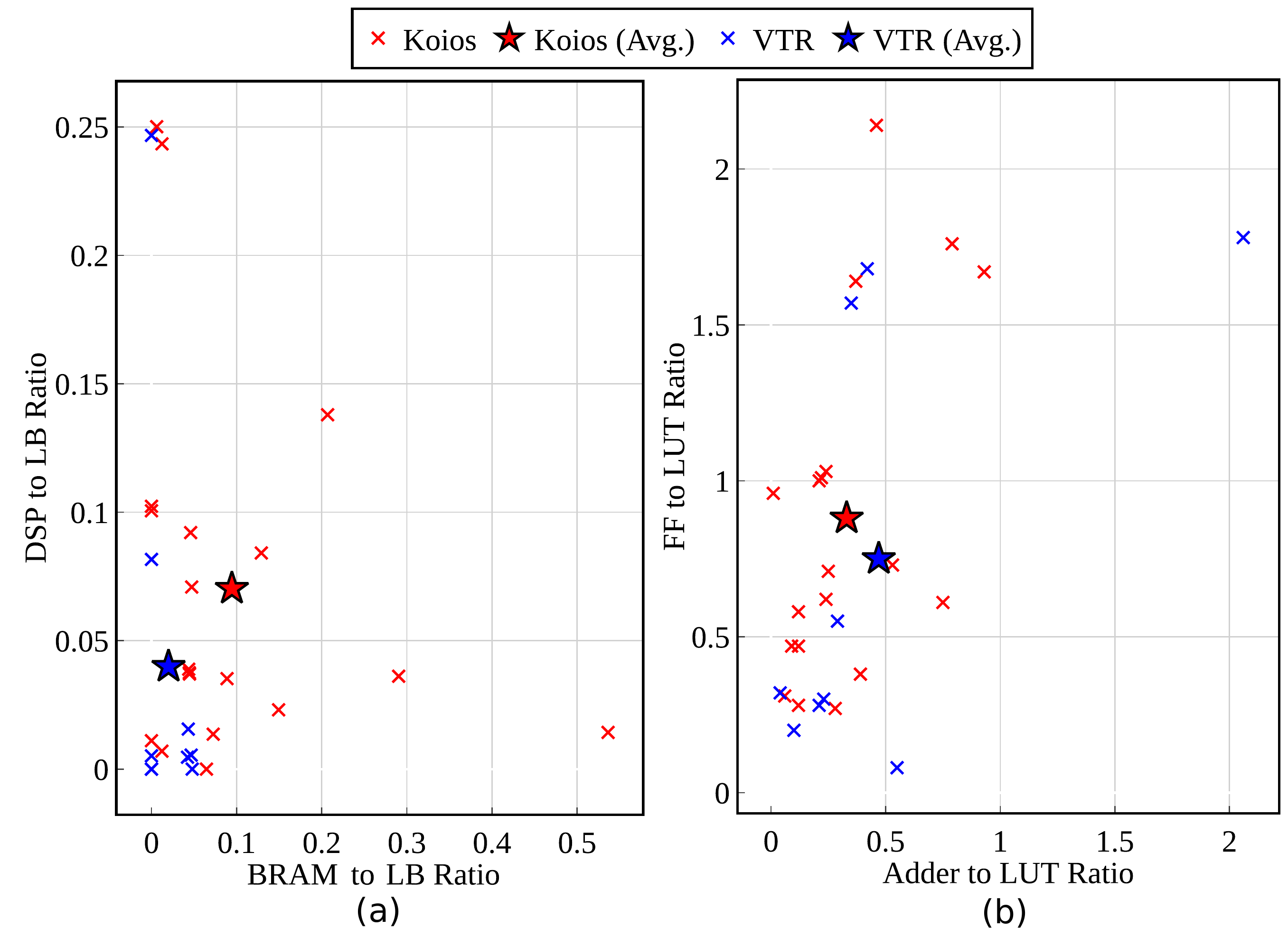}
    \caption{Comparing circuit compositions of \benchname\ \& VTR bench- marks: (a) DSP/BRAM to LB ratios, (b) FF/adder to LUT ratios.}
    \label{fig:scatters}
\end{figure}

Fig. \ref{fig:scatters}a shows a scatter plot of the DSP and BRAM to LB ratios for both \benchname\ (red) and VTR (blue) benchmarks as metrics for their DSP and memory density.
The individual ratios for each of the benchmarks are shown by ($\times$) symbols while the average across the whole benchmark suite is marked by the stars.
The figure shows that, on average, the \benchname\ benchmarks are more DSP and memory rich than the VTR benchmarks.
The \benchname\ suite has a 1.8$\times$ and 4.7$\times$ higher DSP to LB and BRAM to LB ratios, respectively.
The individual benchmarks of the \benchname\ suite are also more scattered and varying across the spectrum of DSP and BRAM compositions.
More importantly, it shows that most of the VTR benchmarks have very low DSP and BRAM densities (except for the only \texttt{stereovision2} outlier circuit), making them inadequate for evaluating any DSP or BRAM architecture modifications.

\begin{figure}[t!]
    \centering
    \includegraphics[width=1\linewidth]{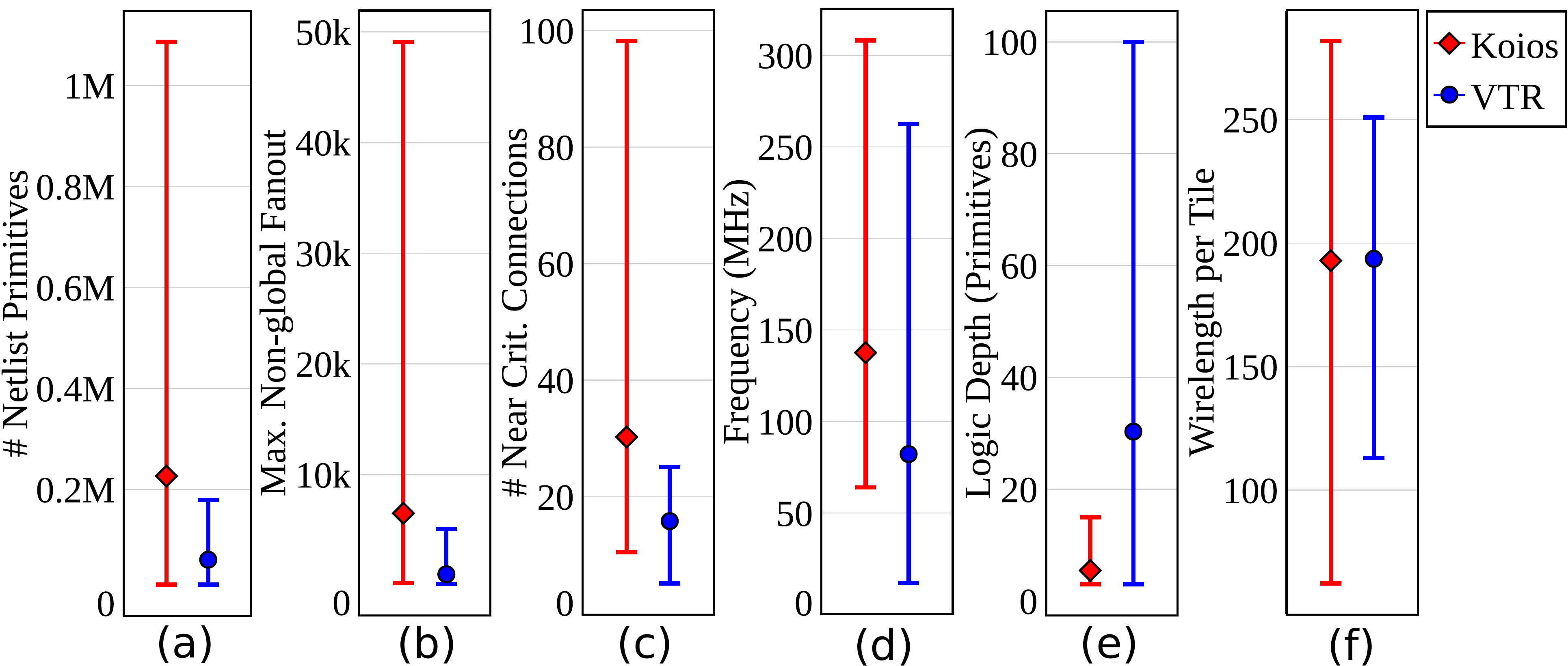}
    \caption{Averages and ranges of key metrics of \benchname\ \& VTR suites.}
    \vspace{-0.1cm}
    \label{fig:ranges}
\end{figure}

Fig. \ref{fig:scatters}b has a similar plot for FF and single-bit adder to LUT ratios.
It shows that the \benchname\ suite has 1.17$\times$ higher ratio between FFs and LUTs which reflects their deeply-pipelined nature, and 30\% lower adder to LUT ratio compared to the VTR suite.
However, the average adder to LUT ratio of the VTR suite is significantly skewed by a single benchmark (\texttt{stereovision2}) which has $60,753$ 1-bit adders and only $29,541$ LUTs.
If we exclude this outlier, the \benchname\ suite has a 1.2$\times$ higher average adder to LUT ratio.

Fig. \ref{fig:ranges} illustrates averages and ranges of key metrics for both \benchname\ and VTR benchmark suites.
Fig. \ref{fig:ranges}a-d show that the \benchname\ benchmarks have 3.7$\times$ more netlist primitives, 6.5$\times$ larger non-global fanouts, 1.9$\times$ more near (top 10\%) critical connections, and 1.7$\times$ higher frequencies on average compared to the VTR benchmarks.
The \benchname\ benchmarks are also scattered across a much wider range of values for each of those metrics.
Fig. \ref{fig:ranges}e shows that the \benchname\ designs have an average of 5 logic levels on the critical path, compared to 30 levels for the VTR benchmarks.
This also reflects the deeply-pipelined nature of our benchmarks which is a key property of modern FPGA designs.
Fig. \ref{fig:ranges}f shows that the two benchmark suites have similar average routed wirelength per tile, with the most wiring dense circuit in \benchname\ having 12\% higher wirelength per tile compared to the most-wiring dense circuit in the VTR suite.
\section{Architecture Exploration Case Studies} \label{section:arch_case_study}

Our \benchname\ benchmark suite is architecture-agnostic and does not depend on any commercial tools for any portion of the FPGA CAD flow.
Thus, it enables the use of these benchmarks to perform flexible FPGA architecture exploration using the fully-open-source VTR flow.
In this section, we perform two example case studies to demonstrate that.

\begin{figure}[t!]
\centering
\includegraphics[width=0.8\linewidth]{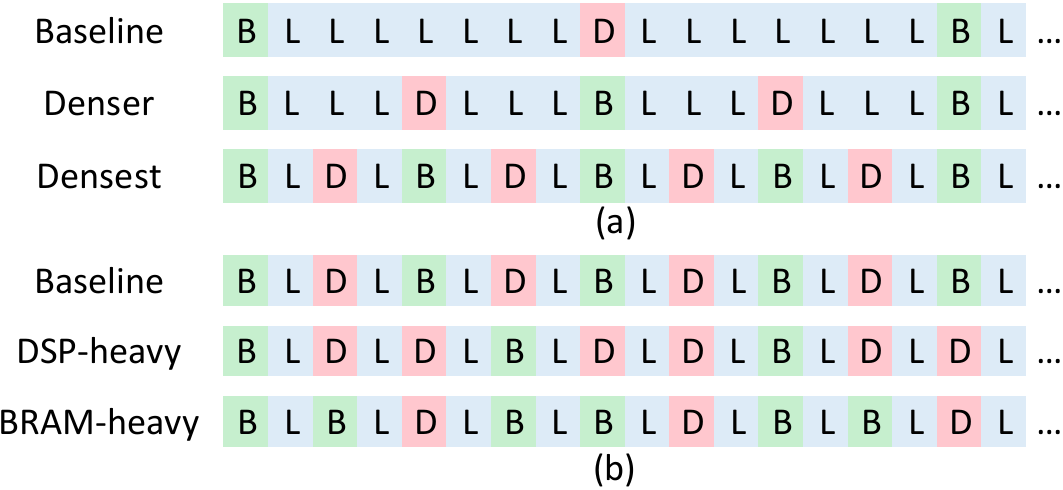}
\caption{FPGA layouts the architectures used in our case studies.}
\label{fig:arch_variations}
\end{figure}

\subsection{Case Study 1: Hard Blocks to Soft Logic Ratio}

\begin{figure}[t!]
\centering
\includegraphics[width=0.7\linewidth]{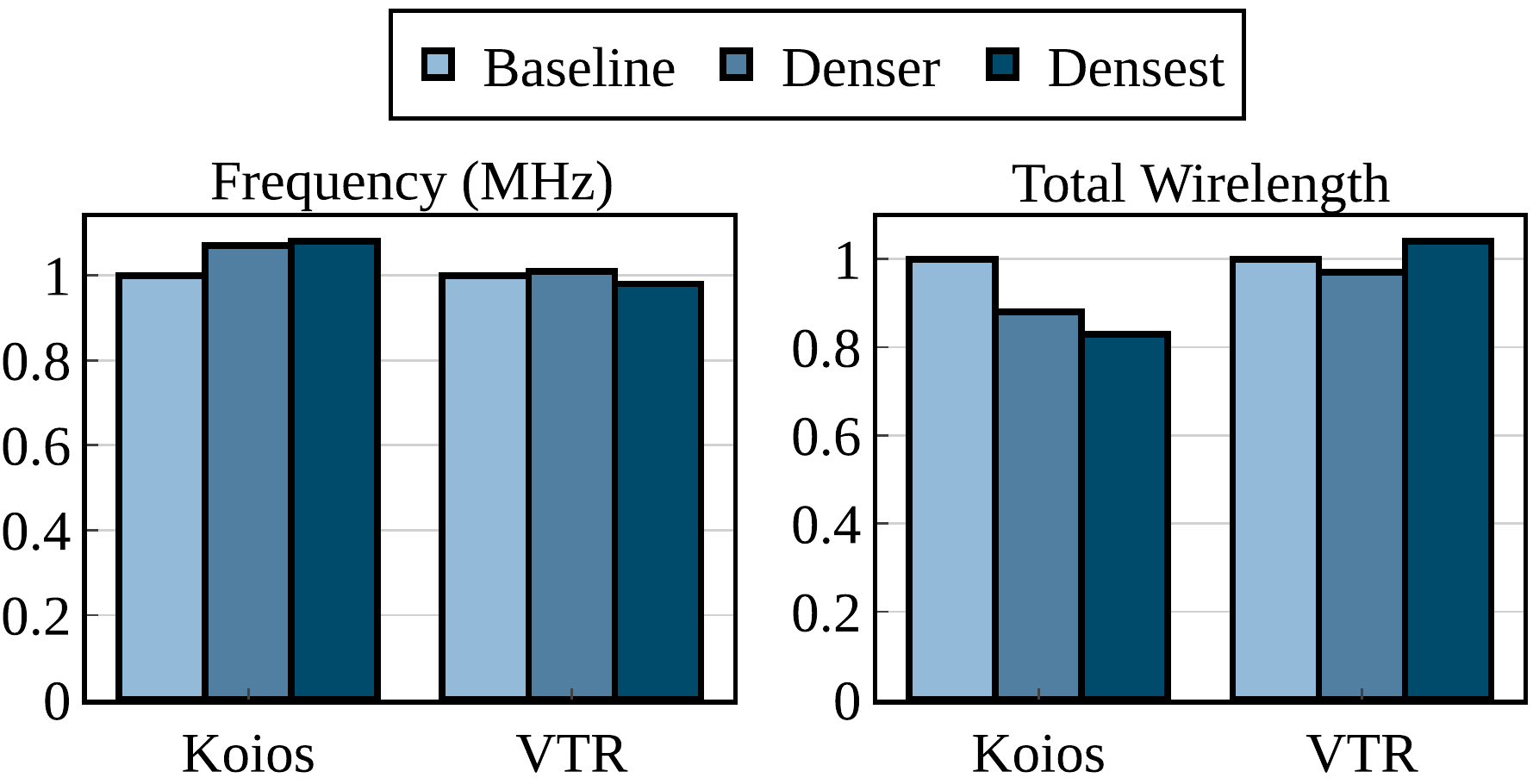}
\caption{Effect of varying the density of DSPs and BRAMs on Koios and VTR benchmark suites.}
\label{fig:ml_vs_non_ml_density}
\end{figure}

As shown in Table \ref{tab:benchmark_properties}, our DL-focused circuits are highly heterogeneous (i.e. DSP and BRAM intensive).
Thus, in our first case study, we vary the density of these hard blocks with respect to soft logic.
We experiment with 3 different density levels, as shown in Fig. \ref{fig:arch_variations}a, with 1:7, 1:3, and 1:1 ratio between hard block and soft logic columns for the baseline, denser, and densest architecture variations, respectively.
We evaluate all three architecture variations using both the \benchname\ and VTR benchmarks.
Fig. \ref{fig:ml_vs_non_ml_density} shows the geomean frequency and total routed wirelength for both suites.
For the DL-oriented \benchname\ benchmarks, the frequency increases and wirelength decreases as the density of hard blocks increases.
Since these benchmarks heavily utilize these blocks, increasing their density in the FPGA grid brings them closer to each other, which in turn reduces the critical paths and total length of used wires.
The densest architecture variation results in 8\% increase in frequency and 17\% reduction in total wirelength on average across all benchmarks in the \benchname\ suite.
For the VTR benchmarks, both frequency and wirelength are slightly improved for the denser variation (1\% higher frequency and 3\% lower wirelength), before getting worse for the densest architecture.
These results show that a higher density of DSPs and BRAMs is favorable for building DL-optimized FPGAs, at the cost of a slight or no degradation in QoR for the general VTR benchmarks (in the densest and denser architecture variations respectively).

\subsection{Case Study 2: DSP to BRAM Ratio}

In our first case study, we varied the ratio of hard blocks to soft logic while keeping a fixed 1:1 DSP to BRAM ratio.
For the second case study, we carry over the best architecture variation for DL benchmarks from the first case study (i.e. densest).
However, we vary the DSP to BRAM ratio between 2:1 and 1:2 to create DSP-heavy and BRAM-heavy variations respectively (in addition to the baseline with 1:1 ratio), as shown in Fig. \ref{fig:arch_variations}b.
Table \ref{tab:num_variation} presents the results of this experiment.
It shows the geomean frequency, routed wirelength, and FPGA grid size for the whole \benchname\ suite, as well as the results for a DSP-intensive benchmark (medium \texttt{tpu\_like} ) and a BRAM-intensive benchmark (\texttt{tiny\_darknet\_like}).
The geomean results do not show a strong trend that clearly favors a specific architecture. 
However, we observe that the DSP-heavy \texttt{tpu\_like} design has 8.6\% higher frequency, 9.3\% lower wirelength, and requires a 25\% smaller chip when implemented on the DSP-heavy architecture compared to the baseline.
It also performs considerably worse on all metrics when implemented on a BRAM-heavy architecture.
Similarly the BRAM-heavy \texttt{tiny\_darknet\_like} benchmark has 7.5\% higher frequency, 6.4\% lower wirelength, and requires a 25\% smaller chip when implemented on the BRAM-heavy architecture compared to the baseline.
These experiments highlight that Koios strikes a good balance between different circuit compositions and can be reliably used for DL-optimized FPGA architecture exploration.

\begin{table}[t!]
  \centering
    \caption{Effect of varying the FPGA's DSP to BRAM ratio.}
    \begin{threeparttable}
    \begin{tabular}{C{0.85cm}@{\hspace{1\tabcolsep}} c@{\hspace{1\tabcolsep}} C{1cm}@{\hspace{1\tabcolsep}} C{1.4cm}@{\hspace{1\tabcolsep}} C{2.1cm}@{\hspace{1\tabcolsep}}}
    \toprule
    \textbf{Metric} & \textbf{Arch.} & \textbf{Geo- mean} & \textbf{DSP-heavy tpu\_like(M)} & \textbf{BRAM-heavy tiny\_darknet\_like} \\
    \toprule
    \multirow{3}{*}{\textbf{Freq.}} & \textbf{Baseline} & 141.2 & 141.1 & 94.1 \\
          & \textbf{DSP-heavy} & 141.9 & \textbf{153.3} & 86.8 \\
          & \textbf{BRAM-heavy} & 140.8 & 120.4 & \textbf{101.1} \\
    \hline
    \multirow{3}{*}{\textbf{WL}} & \textbf{Baseline} & 622,189 & 1,460,366 & 2,076,993 \\
          & \textbf{DSP-heavy} & 623,777 & \textbf{1,325,930} & 2,313,599 \\
          & \textbf{BRAM-heavy} & 641,263 & 1,661,778 & \textbf{1,944,531} \\
    \hline
    \multirow{3}{*}{\textbf{Grid}} & \textbf{Baseline} & 84$\times$84    & 134$\times$134   & 180$\times$180 \\
          & \textbf{DSP-heavy} & 85$\times$85    & \textbf{116$\times$116}   & 220$\times$220 \\
          & \textbf{BRAM-heavy} & 88$\times$88    & 164$\times$164   & \textbf{156$\times$156} \\
    \bottomrule
    \end{tabular}%
    \begin{tablenotes}
    \item Frequency is in MHz, Wirelength (WL) is in units of length 1 wires.
    \end{tablenotes}
    \end{threeparttable}
  \label{tab:num_variation}
\end{table}

\section{Conclusion} 
\label{section:conclusion}

In this paper, we presented \benchname, a DL-focused 
benchmark suite for FPGA architecture and CAD research.
This suite is a diverse collection of 19 curated benchmarks covering various facets of the DL acceleration landscape.
We first introduce the different benchmarks in the suite and highlight their diversity. 
We then present results of running these benchmarks through the VTR flow and compare them to the existing non-DL VTR benchmarks.
Finally, we present two example case studies for DL-optimized FPGA architecture exploration using these benchmarks. 
The \benchname\ suite is open-sourced as a part of VTR and we highly encourage the FPGA community to contribute to this benchmark suite to help build a better and bigger set of DL benchmarks that can guide the design of future FPGA architectures and CAD algorithms.  

\section*{Acknowledgement}
We would like to thank Helen Dai and Zach Zheng from the University of Toronto for contributing to the benchmarks. We are grateful to the National Science Foundation (grant number 1763848), the NSERC/Intel Industrial Research Chair in Programmable Silicon, the Vector Institute for AI, and the Intel/VMWare Crossroads Research Center for funding support. Any opinions, findings, conclusions or recommendations are those of the authors and not of the funding institutions.

\bibliographystyle{IEEEtran}
\bibliography{bibliography}

\end{document}